\documentclass[twocolumn,showpacs,superscriptaddress,amsmath,amssymb]{revtex4}

\usepackage{epsfig,amsmath,amssymb,graphics,color,calc,wasysym}

\newcommand{\be}{\begin{equation}}
\newcommand{\ee}{\end{equation}}
\newcommand{\ba}{\begin{eqnarray}}
\newcommand{\ea}{\end{eqnarray}}

\renewcommand{\phi}{\varphi}
\newcommand{\tw}{t_\mathrm{w}}

\begin{document}

\title{Subdiffusion and intermittent dynamic fluctuations in the aging
regime of concentrated hard spheres}

\author{Djamel El Masri}

\affiliation{Soft Condensed Matter, Debye Institute for NanoMaterials
Science, Utrecht University, Princetonplein 5, 3584 CC,
Utrecht, The Netherlands}

\affiliation{Laboratoire des Collo{\"\i}des, Verres et
Nanomat{\'e}riaux, UMR 5587, Universit{\'e} Montpellier 2 and CNRS,
34095 Montpellier, France}

\author{Ludovic Berthier}
\affiliation{Laboratoire des Collo{\"\i}des, Verres et
Nanomat{\'e}riaux, UMR 5587, Universit{\'e} Montpellier 2 and CNRS,
34095 Montpellier, France}

\author{Luca Cipelletti}
\affiliation{Laboratoire des Collo{\"\i}des, Verres et
Nanomat{\'e}riaux, UMR 5587, Universit{\'e} Montpellier 2 and CNRS,
34095 Montpellier, France}

\date{\today}

\begin{abstract}
We study the nonequilibrium aging dynamics in a system of quasi-hard spheres
at large density by means of computer simulations.
We find that, after a sudden quench to large density,
the relaxation time initially increases exponentially with the
age of the system. After a surprisingly large
crossover time, the system enters the asymptotic aging regime
characterized by a linear increase of the relaxation time
with age.
In this aging regime, single particle motion is strongly
non-Fickian, with a mean-squared displacement increasing
subdiffusively, associated to broad, non-Gaussian tails
in the distribution of particle displacements.
We find that the system ages through temporally intermittent
relaxation events, and a detailed finite size analysis
of these collective dynamic fluctuations reveals
that these events are not spanning the entire system, but
remain spatially localized.
\end{abstract}


\maketitle

\section{Introduction}

Aging refers to the slow evolution with time of
the physical properties of a disordered
material suddenly quenched into a glass phase~\cite{young}.
This might
refer to the evolution of the density of a polymer glass, of
the dielectric response in an organic liquid,
or the height of a gently shaken sandpile. Aging is more
easily detected by focusing on the dynamics of the system,
for instance by measuring correlation or response functions.
One can think of measuring the decay of density fluctuations
using light scattering techniques in colloidal glasses,
or the time-dependent magnetic response in a spin glass.
After several decades of aging studies in glassy materials,
these generic features are very well
documented~\cite{young,leszouches,lucareview}.

Much less is known, however, about the microscopic mechanisms
involved during aging, and how these evolve with time, for
two main reasons. First, the important theoretical developments
from the last decade about aging dynamics mainly stemmed from
`mean-field' types of approaches, which provided detailed
predictions about the behavior of averaged dynamic quantities,
but very little information about microscopic motion~\cite{young}.
Second, it is only relatively recently that the microscopic
dynamics, of, say, molecules in a supercooled liquid, has been
characterized in detail in equilibrium conditions
above the glass transition~\cite{ediger}, and, by comparison, much less
work has been done about the corresponding aging
regime at lower temperatures.

In this paper, we use numerical simulations to study the
aging dynamics of a model system designed to understand the
behavior of dense suspensions of colloidal hard spheres.
Hard spheres represent one of the simplest, and thus most studied,
models to study the glass transition. It is easily
studied in simulations, and the model finds its experimental
realization using well-controlled colloidal suspensions~\cite{naturepusey,ericluca}.
As opposed to molecular liquids, colloidal hard spheres can be studied
at the particle scale using microscopy techniques~\cite{weeks,KegelScience2000}, while dynamic light scattering provides a convenient way to characterize their dynamics in great detail~\cite{martinez,djamel}. It should be noted that while a relatively small set of experiments have been reported on the aging of colloidal hard spheres~\cite{martinez,djamel,SimeonovaPRL2004,weeksaging,weeksaging2}, a much broader literature exists on the out-of-equilibrium dynamics of a variety of more complex colloidal systems. Throughout this paper, we will in particular make reference to findings for colloidal attractive gels~\cite{lucaaging1,DuriEPL2006,DuriPRL2009} and to systems with soft repulsive interactions, such as closely packed soft spheres~\cite{yodh,MazoyerPRL2006,MazoyerPRE2009} or Laponite clay platelets interacting via Coulomb repulsion~\cite{bonn,kaloun,bob1,joshi,lequeux}.

In this work, we will be concerned with three main questions.

(1) {\it How does the structural relaxation slow down with the aging time?}
While numerical simulations of model glasses 
generically find that the structural
relaxation time increases roughly linearly (or sub-linearly) with the
sample age~\cite{kob-barrat},
some light scattering experiments on gels and Laponite report an unexpected
exponential growth of the relaxation time with sample
age~\cite{lucaaging1,bonn,kaloun}, at least at short times.

(2) {\it How do particles move during the aging regime?}
Confocal microscopy experiments report that colloidal particles
move very little in concentrated hard spheres, to the point
that distinguishing thermal vibrations from genuine
relaxation becomes an experimental
challenge~\cite{weeksaging,weeksaging2}. In simulations
as well, particles move very little, leading to
a very slow, typically algebraic, decay of time correlation
functions~\cite{kob-barrat,kob-barrat2,puertas}.
This is in stark contrast with several experimental
reports of sharply decaying time correlation functions in aging
molecular liquids seeded with colloidal particles or in colloidal gels and Laponite, where compressed exponential
decay and ballistic particle motion over large distances were
reported~\cite{lucareview,lucaaging1,bob1,CaronnaPRL2008,bob2}.

(3) {\it How heterogeneous is the dynamics?}
Due to the key role played by dynamic heterogeneity in
equilibrium studies of the glass transition~\cite{ediger}, similar
signatures have been sought in aging materials.
Optical and confocal microscopy studies revealed the existence of rare small-scale
relaxation events involving extended clusters
of particles~\cite{weeks},
whose size is modest and grows very little~\cite{yodh},
or not at all~\cite{weeksaging,weeksaging2},
during aging. This agrees with numerical simulations
on model glasses, where four-point dynamic susceptibility
appear to grow at a very slow rate with the sample age~\cite{parisi,castillo},
suggesting
that if collective displacements occur, they are
correlated over relatively short length scales. This set
of results is however in contrast with another series of observations.
Sudden relaxation events termed (rather dramatically) `earthquakes'~\cite{kob-barrat2}
or `avalanches'~\cite{lee1,lee2} were reported in numerical
simulations of Lennard-Jones glasses and these were suggested
to be spanning the entire system. Similarly, highly
intermittent dynamic fluctuations were experimentally
recorded in several aging systems
from polymer glasses~\cite{buisson} to 
colloidal gels~\cite{lucaaging1,DuriEPL2006,lequeux} and 
soft spheres~\cite{MazoyerPRL2006},
the latter typically
involving ballistic particle motions correlated over very
large distances~\cite{DuriEPL2006,DuriPRL2009,MazoyerPRL2006,MazoyerPRE2009}.

The paper is organized as follows. In Sec.~\ref{model}, we
present the numerical model and techniques
used in the present study. In Sec.~\ref{average},
we describe the average aging dynamics for the
evolution of the energy and relaxation timescale.
In Sec.~\ref{subdiffusion}, we focus on the subdiffusive
and heterogeneous dynamics at the single
particle level. In Sec.~\ref{collective} we present
results concerning the intermittent collective
dynamics of the system. In Sec.~\ref{conclusion}
we discuss our results and conclude the paper with some perspective for
future research.

\section{Numerical model and techniques}
\label{model}

We perform molecular dynamics simulation of
dense systems of strongly repulsive particles
interacting with a very steep pair potential
designed to model the behavior of colloidal
hard spheres~\cite{Voigtmann}:
\begin{equation}
\label{potential}
V(r_{ij})=\epsilon \left( \dfrac{\sigma_{ij}}{r_{ij}} \right)^{36},
\end{equation}
where $\epsilon$ is an energy scale,
$r_{ij} = |{\bf r}_i - {\bf r}_j|$, ${\bf r}_i$ being the position
of particle $i$, and $\sigma_{ij} \equiv (\sigma_i +\sigma_j)/2$,
where $\sigma_i$ represents the diameter of particle $i$.
To prevent crystallization occurring in
dense systems of hard spheres,
we introduce a size polydispersity and
draw the particle diameters from a
flat size distribution, $\sigma_i / \sigma \in [0.8, 1.2]$, so that
the average diameter is $\overline{\sigma_i} = \sigma$,
and the polydispersity $\delta$ is given by
\be
\delta = \sqrt{ \frac{\overline{\sigma_i^2}}{\overline{\sigma_i}^2}
- 1 } \approx 11.5 \%.
\ee

To study the dynamics of the system we solve
Newton's equations for a system composed of $N$ particles,
using a velocity Verlet algorithm~\cite{allen} 
in a cubic box of linear size
$L$. We use periodic boundary conditions in the three
directions of space. We have studied two system sizes,
$N=4000$ and $N=500$. When dealing with averages, we shall
report results for the largest system size, while a
comparison between the two systems will allow us to perform a detailed
comparison of the dynamic fluctuations occurring
during the aging dynamics as a function of system size.
We perform simulations in the $NVT$ ensemble, and we control
the temperature by rescaling
velocities every 100 molecular dynamics timesteps
in order to maintain the kinetic energy to the desired constant
value.

For the inverse power law
potential in Eq.~(\ref{potential}), density and temperature cannot
be controlled independently, as rescaling the density by a factor
$\lambda$ is equivalent to rescaling the temperature
by a factor $\lambda^{-1/12}$. Thus we fix the temperature and
energy scales, $k_B T = \epsilon = 1/3$, and simply
vary the density, $\rho = N/L^3$, of the system~\cite{Voigtmann}.
To ease the comparison with hard sphere experiments, we express
our results using the volume fraction, $\phi$, rather than density
$\rho$ , where
\be
\phi = \frac{\pi \rho}{6 N} \sum_i \sigma_i^3.
\ee
In the following, we express
length scales in units of $\sigma$,
and timescales in
units of $\tau_0 = \sigma \sqrt{m /\epsilon}$,
where $m$ is the mass of the particles. We
use a time discretization $\Delta t = 0.01 \tau_0$,
which ensures a proper integration of the equations of motion.

We shall study the aging dynamics of samples
in the range $\phi = 0.553 - 0.662$. To obtain
reproducible results, we need to produce fully disordered initial states.
To this end, we
first equilibrate the system at very low volume fraction,
$\varphi = 0.14$. We then compress the system very rapidly
to the desired final volume fraction. The compression is done in
small successive steps in order to avoid
large overlaps leading to very large repulsive forces.
The age $\tw$ of the system is counted from the time when
the system reaches the final volume fraction.
To increase the statistical significance of our results,
we have performed 5 independent runs at each volume fraction
with $N=4000$ particles, starting from independent
configurations compressed from the fluid at $\varphi=0.14$.
During the course of the simulations we found no
sign of incipient crystallization in the system, while
crystallization was a major obstacle in an initial set of studies
using a smaller polydispersity, $\delta \approx 6~\%$,
for which we do not present results.

We have previously studied the
equilibrium dynamics of this system~\cite{djamel}. We found that
the dynamics slows down considerably when $\phi$ increases
above $\phi \approx 0.50$. Fitting the increase of the
relaxation to a power-law divergence at some critical
volume fraction $\phi_c$, we obtained $\phi_c \approx 0.592$.
This fit then locates the (apparent) mode-coupling singularity
for this system~\cite{brambilla}, which should serve as a
reference volume fraction for the aging studies below, since
it becomes very difficult to reach thermal equilibrium
within the accessible numerical timescales when $\phi$ increases beyond $\phi_c$.

\section{Towards the asymptotic aging regime}
\label{average}

\begin{figure}
\center
\includegraphics[width=0.95\linewidth]{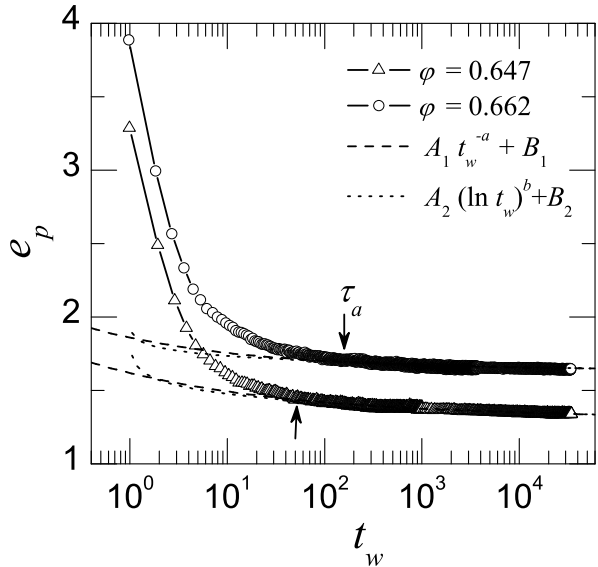}
\caption{Aging of the potential energy density,
Eq.~(\ref{eqep}), for
$\varphi=0.647$ and $\varphi=0.662$. Data are equally well fitted with
algebraic, $A_{1}\times \tw^{-a} + B_{1}$,
or logarithmic, $A_{2}\times \left( \ln \tw \right)^{b}
+ B_{2}$, time dependences for waiting times
larger than a surprisingly large crossover timescale, $\tau_a$,
indicated by arrows.}
\label{Potenergy}
\end{figure}

We now present the results of our study, starting from the
time evolution of the simplest quantity we can monitor in our study,
the potential energy density, defined as
\be
e_p(\tw) = \left\langle \frac{1}{N}
\sum_{i=1}^N \sum_{j>i} V(r_{ij}) \right\rangle,
\label{eqep}
\ee
where the brackets stand for an average over independent
initial configurations. We present representative results
for the time evolution of $e_p(\tw)$ at large densities
in Fig.~\ref{Potenergy}.

The central observation from this figure is that,
at these large densities, the potential energy keeps evolving over
the 6 decades of the simulation and never reaches
an asymptotic plateau. This simply confirms that thermal
equilibrium cannot be reached at these volume fractions, because
the equilibrium relaxation time is much larger than the maximum time
accessible to our
simulations. More in detail, we also observe that the energy
evolves initially quite rapidly, decreasing by a factor of about 2
when time increases from $\tw =1$ to $\tw = 10^2$, while it
evolves only by a few percent when time further increases by
two additional decades. Thus, we clearly see the effect of `aging',
since dynamical evolution slows down considerably as the
age of the system increases, as is well-known from decades
of experimental aging studies in many different materials.

To describe the slow evolution of the potential energy,
we fitted our data to both a power law decay,
$e_p(\tw)  = A_{1}  \tw^{-a} + B_{1}$, and
a logarithmically slow evolution,
$e_p(\tw) =  A_{2} \left( \ln \tw \right)^{b} + B_{2}$.
We find that both fits give an equally good description of
the final, slow evolution of the potential energy.
The fit parameters, $a \approx 0.2-0.3$ and $b \approx 0.1$
confirm that the evolution of the energy is indeed extremely slow.

We also find that the fits only hold when $\tw$ is very large:
the first few decades of the simulation, corresponding to a faster
evolution of $e_p(\tw)$ are not well described by this asymptotic slow
decay. This implies that a relatively long time is needed
for the system to enter the asymptotic aging regime.
Moreover, we find that the time $\tau_a$
required to enter the asymptotic regime
increases when density increases and the system is quenched deeper
into the glassy state. This implies that considerable
care must be taken when performing data analysis of the aging
regime, since it already takes a large part of the simulation
simply to enter the final regime, indicated by the arrows
in Fig.~\ref{Potenergy}. It is likely that the first regime
corresponds to faster processes where the largest overlaps
created during the compression are removed, producing
heat which is then removed by our thermostatting procedure.
Thus, no `universal' characteristics are to be expected in this
time regime, which might well depend quite strongly on the details
of the preparation procedure, or the chosen thermostat.
Note that $\tau_a \approx 10^2 \gg \tau_0$ for the data presented in
Fig.~\ref{Potenergy}, meaning that scaling behavior
might only be observed in the demanding limit of
\be
\tw \gg \tau_a \gg \tau_0.
\label{taua}
\ee

\begin{figure}
\center
\includegraphics[width=0.95\linewidth]{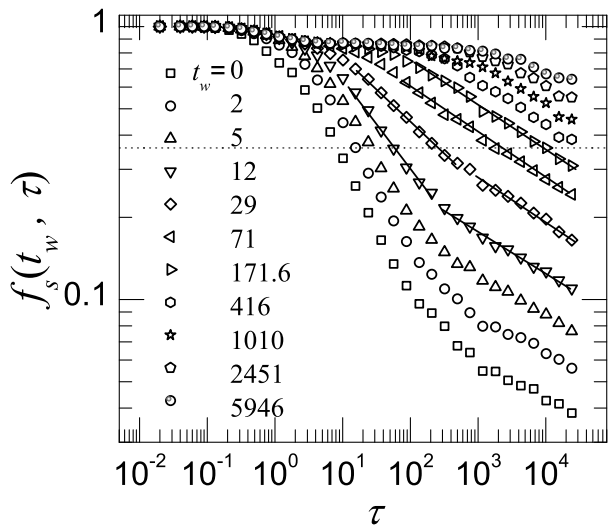}
\caption{Self-intermediate scattering function,
Eq.~(\ref{fself}), for $\varphi=0.647$ in a log-log representation.
The long-time decay is fitted with
two power law decays for $\tw < \tau_a$, a single one
for $\tw > \tau_a$, as shown with full lines.
The exponent for the latter is nearly constant, $\approx -0.15$.
The horizontal dashed line indicates $f_s = 1/e$. }
\label{Fs-phi0.647}
\end{figure}

It is crucial to recognize the existence of two distinct
aging regimes in order to analyze properly the scaling
properties of dynamic functions in the asymptotic aging regime,
as we find that these two distinct regimes in fact
affect most of the measurements we made in our simulations.
In Fig.~\ref{Fs-phi0.647} we show the behavior of the
self-intermediate scattering function following
a quench at $\phi=0.647$. This two-time quantity is defined as
\begin{equation}
f_s(\tw,\tau) =  \left\langle \frac{1}{N}
\sum_{j=1}^N
e^{i
{\bf q} \cdot \left [{\bf r}_j(\tw+\tau)-{\bf r}_j(\tw)\right ]}
\right\rangle,
\label{fself}
\end{equation}
and we perform measurements at $q = 7.8$, close to the first peak
in the static structure factor.
As is well-known, two-time quantities reveal the non-stationary
evolution of the system in a very direct manner. In agreement
with previous work~\cite{kob-barrat,puertas},
we find that the decay with the delay time $\tau$
of $f_s(\tw,\tau)$ occurs in a two-step manner, the slow decay
being a strong growing function of the waiting time $\tw$. The faster initial decay is much less
sensitive to the waiting time, and corresponds physically
to particle vibrations in a nearly frozen amorphous structure.

The existence of two distinct aging regimes is obvious
from these data, since the slow decay of $f_s$ clearly shows
a crossover for time delay $\tau$ such that
$\tau + \tw \approx \tau_a$, as shown in Fig.~\ref{Fs-phi0.647},
while the data become simpler to describe when
$\tw > \tau_a$ as the entire decay then takes place
in the asymptotic long-time regime
defined by Eq.~(\ref{taua}).
As found in other systems~\cite{kob-barrat2}, we find that the long-time
decay of the self-intermediate scattering function
is well described, in the aging regime, by a power law,
$f_s(\tw,\tau) \sim \tau^{-\alpha}$, as indicated in
Fig.~\ref{Fs-phi0.647} by the full lines.
As noted before, such an algebraic decay is in contrast with
equilibrium measurements at lower volume fraction
which typically display stretched exponential relaxations.
Stretched exponential decays are only seen for shallow quenches 
when a crossover towards thermal equilibrium is possible.
The small value of the exponent, $\alpha \approx 0.15$ means
that relaxation occurs over a very broad time window,
and that it is not possible to define a `typical'
relaxation time for the system.

\begin{figure}
\center
\includegraphics[width=0.95\linewidth]{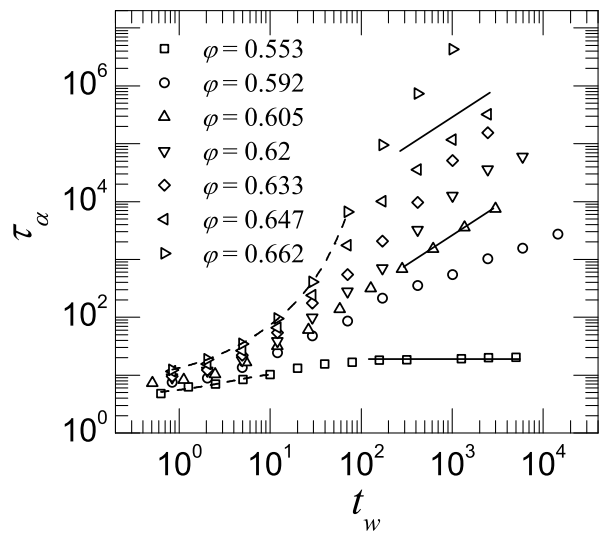}
\caption{Evolution of the relaxation time, $\tau_{\alpha}(t_w)$,
for different volume fractions.
The system at $\varphi=0.553$ reaches equilibrium
(horizontal line). A simple aging regime, $\tau_{\alpha} \sim \tw$,
indicated by the full lines, is reached
for larger densities, after a transient regime where
$\tau_\alpha$ grows exponentially with $\tw$, indicated
with a dashed line which ends for $\tw \approx \tau_a$.}
\label{tautw}
\end{figure}

Nevertheless, to quantify the aging of the dynamics, we follow
the common practice and define an $\alpha$-relaxation time
$\tau_{\alpha}(\tw)$ from the time decay of $f_s$
as $f_s(\tw , \tau_\alpha) = 1/e$, as indicated
by the horizontal dashed line in Fig.~\ref{Fs-phi0.647}.
In Fig.~\ref{tautw}, we report
the evolution of $\tau_{\alpha}(t_w)$
with waiting time for all volume fractions investigated in
this work, from $\phi=0.553$ to $\phi = 0.662$.
Again, we observe two distinct regimes for the evolution of
$\tau_{\alpha}(t_w)$. While for $\tw < \tau_a$,
$\tau_{\alpha}$ seems to increase exponentially
with waiting time,
\be
\tau_\alpha \sim \exp (c \tw),   \quad \tw \ll \tau_a,
\label{exp}
\ee
with $c$ a numerical constant,
its growth is better described
by an algebraic law at long times, $\tw > \tau_a$,
which is well compatible with a so-called `simple aging'
behavior~\cite{young},
\be
\tau_\alpha \sim \tw, \quad \tw \gg \tau_a.
\label{simple}
\ee
This simple aging behavior is shown with full lines
in Fig.~\ref{tautw}. In agreement with the observations
in Fig.~\ref{Potenergy}, we find that it takes an increasingly
long time for the system to enter the asymptotic (simple) aging
regime when $\phi$ increases. For the larger $\phi$ studied,
$\phi = 0.662$, the simple aging regime is not reached during the
course of our simulations, and the system appears to be
in the crossover between (\ref{exp}) and (\ref{simple}), and the
system seems to undergo an effective `super-aging', i.e. its
relaxation time increases faster than its age.
Finally, the opposite `sub-aging' behavior is found when
volume fraction is not too large, for instance $\phi = 0.592$,
because the system is crossing over towards thermal equilibrium,
such that $\tau_\alpha$ should saturate at long waiting times
to the equilibrium relaxation time.
These observations show that a complex behavior of $\tau_\alpha$,
as often reported in numerical works~\cite{kob-barrat,lee3} and experiments on colloidal gels and Laponite~\cite{lucaaging1,bonn,kaloun}, may be
due to the occurrence of multiple crossovers which are highly
sensitive to volume fraction.

These observations are also in agreement with the common
observation of sub-aging behavior in aging molecular liquids,
for which experiments are traditionally performed not very
far below the glass temperature~\cite{young,struik}. Exponential growth
of the relaxation time sometimes (but not always)
followed by algebraic growth, has been reported in a few experiments
on colloidal glasses or gels as well~\cite{lucaaging1,bonn,kaloun}.
Our results thus suggest
that this exponential growth might well be a transient behavior,
which can persist, however, over a very large time window,
in particular for very deep quenches. Our results
also highlight the difficulty in analyzing the scaling properties
of two-time dynamic quantities in numerical simulations, since
the asymptotic aging regime is only accessed after
a time $\tau_a$ which can be very large, thus
decreasing the window where `universal' aging properties
can be probed. This might well explain some conflicting
results and analysis reported in the literature
regarding, e.g., the proper scaling behavior of the self
intermediate scattering function in Lennard-Jones
glasses~\cite{kob-barrat3,rieger}.

\section{Subdiffusive and heterogeneous dynamics}
\label{subdiffusion}

The above findings that the energy decays very slowly, while
time correlation functions decay in an algebraic manner
indicate that there is actually very little dynamics
taking place in the system.
To confirm this idea, we measured the averaged
mean-squared displacement,
\begin{equation}
\Delta r^{2}(\tw, \tau) = \left\langle
\frac{1}{N} \sum_{i=1}^N
|{\bf r}_{i}(\tw+\tau)-{\bf r}_{i}(\tw)|^{2}\right\rangle.
\label{MSD}
\end{equation}

\begin{figure}
\center
\includegraphics[width=0.95\linewidth]{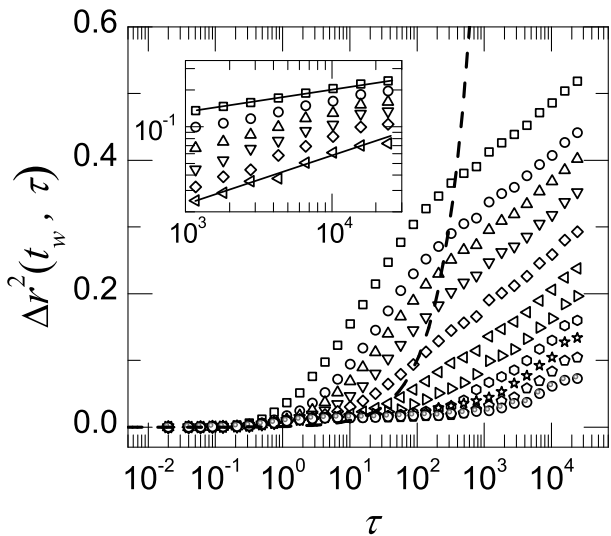}
\caption{Mean-squared displacements, Eq.~(\ref{MSD}),
at $\varphi=0.647$ with symbols as in
Fig.~\ref{Fs-phi0.647}. The dashed line indicates diffusive
behavior. Inset: log-log plot of the data for $\tw > \tau_a$
and large $\tau$ with subdiffusive fits,
Eq.~(\ref{eqnu}), indicated with full lines.}
\label{MSDfig}
\end{figure}

In Fig.~\ref{MSDfig}, we present the time evolution
of  the mean-squared displacements at volume fraction
$\varphi=0.647$ measured at different waiting times, using
both a standard log-log representation (inset) and a lin-log
representation (main plot). As found in Fig.~\ref{Fs-phi0.647}
for the self-intermediate scattering function, the dynamics
proceeds again in a two-step fashion with a rapid, ballistic
regime at short-time, corresponding to vibrations
of the particles in a frozen amorphous structure,
followed by a much slower, waiting time dependent
structural relaxation, which becomes slower when $\tw$ increases.
This implies that the aging of the system corresponds,
at the microscopic scale, to a dramatic slowing down
of single particle displacements, as found in
experiments~\cite{weeksaging,yodh,SimeonovaPRL2004}.

It is significant that in order to represent graphically
the behavior of the mean-squared displacements, we had to use a linear
scale for its amplitude, with a range which remains smaller than
$\Delta r^2 = 0.5$. This implies that over
the entire duration of the simulation,
particles, on average, move a distance which is smaller
than the mean particle diameter: despite the non-trivial aging
dynamics we discuss, it should be clear that the structure of the material
remains in fact almost frozen as soon as the system is quenched in the
glassy phase, with essentially no large-scale dynamics taking place.

The linear representation in  Fig.~\ref{MSDfig} also confirms
the existence of the two distinct aging regimes discussed above,
thus directly revealing the influence on the dynamics at the particle scale of the
timescale $\tau_a$.
Again, the data obtained for $\tw < \tau_a$ present
a crossover for a delay $\tau$ given by $\tau+\tw \approx \tau_a$,
as seen in Fig.~\ref{MSDfig}. Thus, in the following
we again focus on waiting times that are larger than $\tau_a$,
which, we believe,
reflect better the `universal', quench-independent
nature of the aging regime in concentrated hard spheres.

In the long time regime, the dashed line in Fig.~\ref{MSDfig},
which corresponds to a diffusive growth, $\Delta r^2 \sim \tau$,
is obviously an incorrect description of our data since
the displacements seem to increase much more slowly than
diffusively. As shown in the inset, a subdiffusive
law describes our results very satisfactorily,
\be
\Delta r^{2}(\tw,\tau) \sim \tau^{\nu}, \quad \tw \gg \tau_a,
\label{eqnu}
\ee
where $\nu < 1$ is an exponent characterizing the subdiffusion.

\begin{figure}
\center
\includegraphics[width=0.95\linewidth]{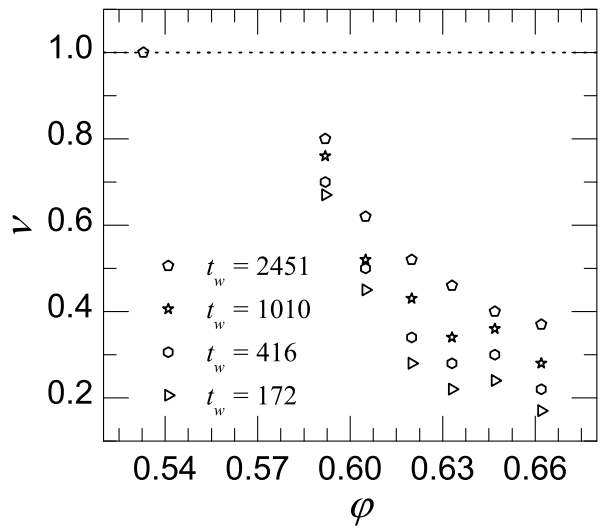}
\caption{Evolution of the subdiffusive exponent
$\nu$ for different $\tw$ and $\phi$. The exponent is further from
its equilibrium diffusive limit $\nu=1$
for smaller $\tw$ and larger $\phi$.}
\label{Sub-diff}
\end{figure}

We have measured $\nu$ over a broad range of waiting times
and volume fractions, and we present its evolution
in  Fig.~\ref{Sub-diff}. Apart from the low volume fractions
where (diffusive) thermal equilibrium is reached rapidly, we systematically
find that the system obeys subdiffusive rather than diffusive
behavior. At a given volume fraction, the system
gets closer to equilibrium for larger
$\tw$, and correspondingly we find that the exponent $\nu$ increases
with the age of the system, although it always remains very far
from its equilibrium, diffusive value $\nu = 1$.
Note that an increasing $\nu$ does not imply that particles
move faster at large waiting times, as the prefactor
in the power law (\ref{eqnu}) is itself a decreasing
function of $\tw$, which implies that the total
amplitude of the particle displacements indeed decreases
with the age of the system.

As volume fraction increases, the system is quenched
deeper and deeper into its glass phase, and deviations
from diffusive behavior are correspondingly larger, which
translates into a subdiffusive exponent $\nu$ which gets
smaller when $\phi$ increases, see Fig.~\ref{Sub-diff}.
This simply means that particles
move less and less when $\phi$ becomes larger, which is physically
expected.

While the mean-squared displacement quantifies
the average dynamical behavior of the system,
it does not convey much information on dynamic fluctuations, which
are recognized as an important feature of the single
particle dynamics in glassy materials.
To quantify this `dynamical heterogeneity' during aging,
we measure the probability distribution function of the single particle displacement,
also known as the self-part of the van-Hove correlation function:
\begin{equation}
\label{vanHove}
G_s(x, \tw, \tau)= \left\langle \frac{1}{N} \sum_{i=1}^N
\delta \left[ x -  ({x}_i(t_w+\tau)-{x}_i(t_w) )
\right] \right\rangle,
\end{equation}
where $x_i(t)$ represents the projection of ${\bf r}_i(t)$ along
the $x$-axis. To improve the statistics, we use
the isotropy of the system and further average
$G_s(x,\tw,\tau)$ along the three directions of space. For simplicity, we keep the notation $G_s(x, \tw, \tau)$ for the van-Hove function averaged over the $x$, $y$ , and $z$ directions.

\begin{figure}
\center
\includegraphics[width=0.95\linewidth]{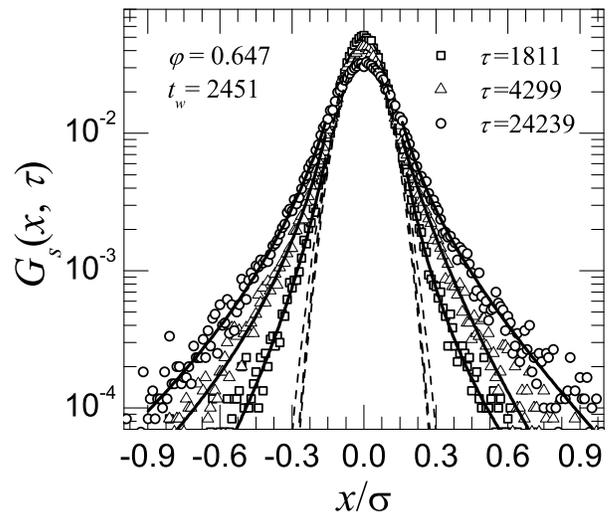}
\caption{Distribution of single particle displacements,
Eq.~(\ref{vanHove}), for $\phi=0.647$, $\tw = 2451$,
and various $\tau$ indicated in the figure.
While a Gaussian (dashed lines) fits the center of the distribution,
the tails are broader and well described by Eq.~(\ref{tail}),
as shown by the full lines.}
\label{van-Hove}
\end{figure}

To gain deeper insight into the subdiffusive behavior
described above, we present in Fig.~\ref{van-Hove} the typical
shape and evolution of the van-Hove function measured
for $\phi=0.647$, for a large waiting time, $\tw = 2451$
in the asymptotic aging regime, and a delay $\tau$
within the long-time subdiffusive regime. The data are presented
in a lin-log scale, where Fickian behavior leading to a Gaussian
shape of the particle distribution would appear as an inverted
parabola, as shown by the dashed lines.
These data are highly reminiscent of the equilibrium
findings that van-Hove functions are well-described
at short displacements by a Gaussian form,
with tails that are much broader than the Gaussian prediction~\cite{weeks,KegelScience2000}.
This implies
that a small fraction of the particles move significantly farther than
what is expected from a purely diffusive and homogeneous process.
This is the most basic observation that the system
is dynamically heterogeneous and can be described,
as being composed of distinct families of `slow'
and `fast' particles, a well-established distinctive
feature of many disordered, glassy materials~\cite{ediger}.

How should we describe the functional form of
the tails of these distributions?
At thermal equilibrium, it is now understood that the
tails are well-described by an exponential (rather than
Gaussian) decay, $G_s(x,\tau) \sim \exp(-|x| / \lambda)$,
for $\tau$ corresponding to the alpha-relaxation timescale~\cite{pinaki}.
The physical origin of the exponential, detailed in
Refs.~\cite{pinaki,epl,pinaki2}, is due to the stochastic nature of
the diffusion process in disordered materials, which
is well described by the
formalism of continuous time random walks (CTRW)~\cite{MW}.
In this description, particles in a dense supercooled liquid
undergo a succession of long periods of vibrations within the
cages formed by their neighbors, followed by rapid
`jumps'. Both the size of the jumps and, more importantly,
the timescale separating them are statistically distributed quantities.
The existence of these statistical distributions
directly accounts for the exponential form of the
tails in the van-Hove functions~\cite{pinaki}.

The natural extension of these considerations
to the non-equilibrium aging regime studied
in the present work is the aging continuous time random walk
(ACTRW) formalism~\cite{actrw},
whose main features are those of the CTRW recalled above. The only difference
lies in the functional form used for the distribution
of times separating the jumps, which acquires `fat',
non-normalizable tails in the aging regime, in
direct analogy with the trap model introduced
by Bouchaud to describe aging dynamics in glasses~\cite{jptrap}.
In this approach, the tails of the jump time distribution
decay as $\psi(t) \sim t^{-1-\nu}$, where $\nu<1$
is the subdiffusion exponent introduced
in Eq.~(\ref{eqnu}), while the tails of the self-part of
the van-Hove function are not exponential anymore,
but are instead asymptotically described by~\cite{metzler}
\be
G_s (x, \tw, \tau) \sim |x|^{-\gamma} \exp \left( -(|x| /\lambda)^{\beta}
\right),
\label{tail}
\ee
where the exponent $\beta$ is also connected to
$\nu$ via the relation
\begin{equation}
\beta= 2 / (2-\nu),
\label{eqbeta}
\end{equation}
so that $\beta=2$ (Gaussian) is recovered
when $\nu = 1$ (Fickian diffusion). For subdiffusive
processes, $0 < \nu < 1$, one expects
instead $1 < \beta < 2$.

To test in detail the ACTRW picture, one should
measure the distribution of jump times $\psi(t)$, and use it to
directly predict the mean-squared displacements and
van-Hove functions~\cite{rottler}. This is an interesting project, but is
beyond the scope of the present study. Instead, we more simply use
Eq.~(\ref{tail}) as a theoretically motivated fitting formula
for the tails of the $G_s$ distributions. As shown in Fig.~\ref{van-Hove},
such a fit describes the tails very well. We find
a similarly good description for a broad range of
volume fractions and timescales.
Unfortunately our data is not accurate enough to allow
for a very precise determination of all the fitting
parameters. Thus, the following results should be discussed
at a qualitative, rather than quantitative, level.

\begin{figure}
\center
\includegraphics[width=0.95\linewidth]{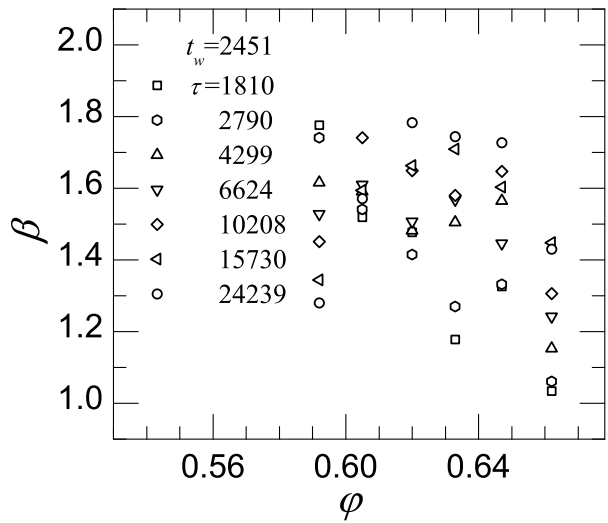}
\caption{Evolution of the exponent $\beta$ obtained
by fitting the tails of the distribution
of particle displacements to Eq.~(\ref{tail}),
for $\tw=2451$ and several $\tau$ and $\phi$.}

\label{Beta}
\end{figure}

In Fig.~\ref{Beta} we present the evolution of the
exponent $\beta$ describing the tails of the van-Hove functions
in the asymptotic subdiffusive aging regime. We remark that these
data have considerably more scatter than the data for the subdiffusive
exponent $\nu$, confirming the difficulty to obtain a reliable determination
of $\beta$ from our data. Regardless of the noise, we note that
$\beta$ values are in the interval $\beta \in [1,2]$, as expected from Eq.~(\ref{eqbeta}), suggesting that our description
is physically sound. Moreover, the data at large density
present a systematic evolution of $\beta$, which increases towards 2
with $\tw$. Assuming that Eq.~(\ref{eqbeta}) is correct, this would imply that
$\nu$ increases towards unity with $\tw$, as indeed is observed in Fig.~\ref{Sub-diff}. The data at moderate
volume fractions, $\phi<0.60$,
are more difficult to interpret as $\beta$ seems
to decrease with $\tw$ in this regime,
while $\nu$ was found to increase. This could be due to the
fact that for $\phi = 0.592 \approx \phi_c$, the subdiffusive aging regime
is crossing over towards equilibrium during the simulation,
and so the results could be a subtle combination
of the exponential tail reported at intermediate times for
equilibrium dynamics~\cite{pinaki}, and the subdiffusive regime
found deeper in the glass and described in the present study.

\section{Intermittent dynamic fluctuations}
\label{collective}

Our study of single particle dynamics suggests that
particles undergo very little dynamics, leading to
a subdiffusive growth of mean-squared displacements,
associated to a broad distribution of particle
displacements. The picture of ACTRW which we used to describe
our data is very similar in spirit
to Bouchaud's trap model, and both suggest that the aging
dynamics of concentrated hard spheres is a temporally
intermittent process~\cite{monthus}. It is the aim of this section
to establish whether this picture is indeed correct.

\begin{figure}
\center
\includegraphics[width=0.95\linewidth]{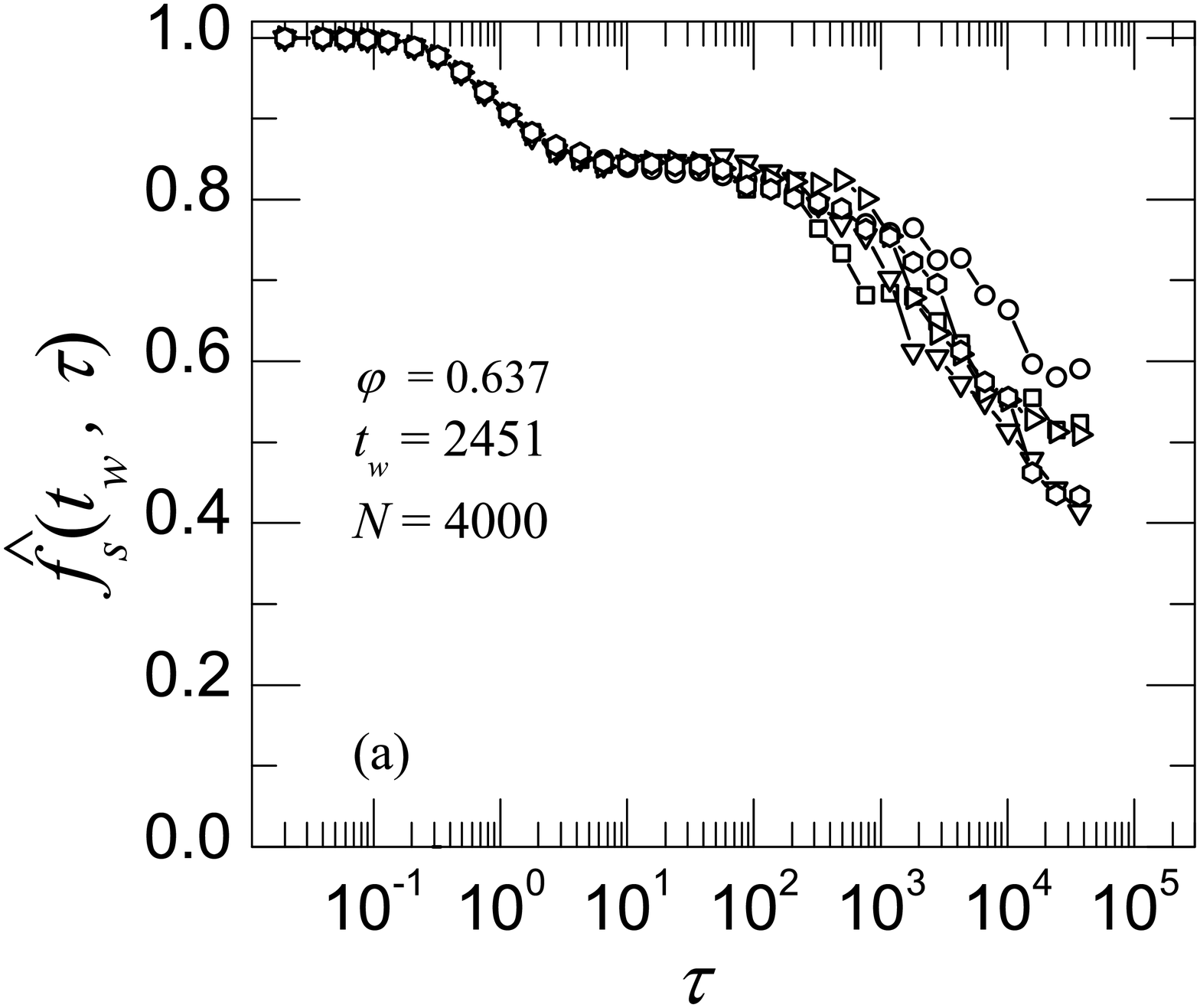}
\includegraphics[width=0.95\linewidth]{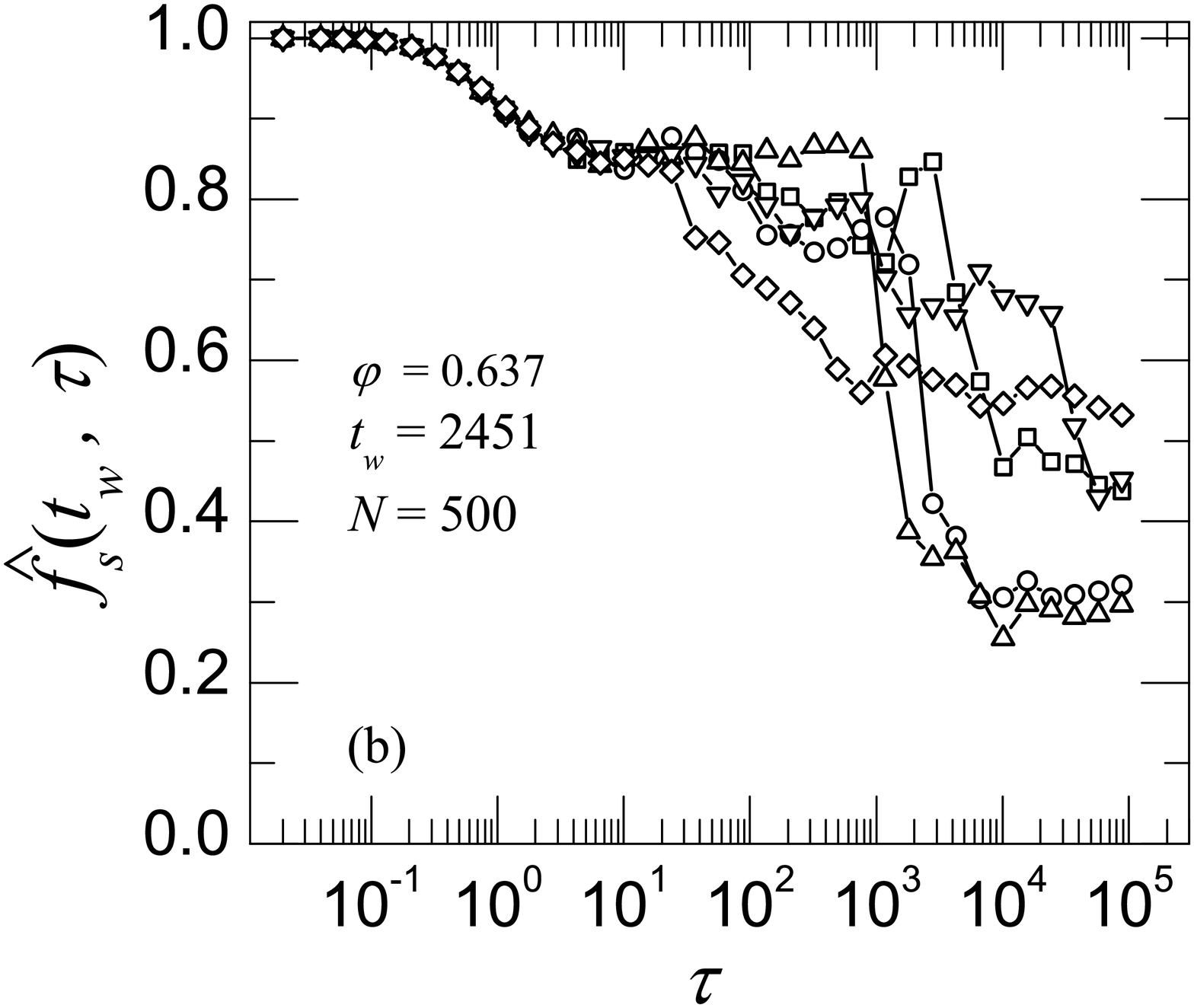}
\caption{Instantaneous (unaveraged)
intermediate scattering function, $\tilde{f_s} (\tw = 2451, \tau)$ for
the system at $\phi=0.637$ and 5 independent quenches.
(a) measured in systems with $N=4000$;
(b) $N=500$. While small run-to-run fluctuations are observed
in (a), a wide variety of
sudden decorrelation events are observed in (b).}
\label{Fs-Earthquaks-N500-N4000}
\end{figure}

To this end, we must turn to collective
dynamic observables, and resolve the aging dynamics
in space and time, and focus on $\tilde{f_s}$, the instantaneous
(unaveraged) value of $f_s$. We present in
Fig.~\ref{Fs-Earthquaks-N500-N4000}(a) the results of
independent realizations
of the dynamics at a given waiting time, $\tw=2451$,
and volume fraction, $\phi=0.637$. For this system, containing
$N=4000$ particles, we observe small run-to-run fluctuations,
which show that resolving the temporal evolution of the dynamics
in a system of linear size $L \sim 15 \sigma$ is not sufficient to reveal
significant dynamic fluctuations.

Thus, we improve our spatial resolution
and repeat this analysis for a smaller system with
$L \sim 7.5 \sigma$ which contains $N=500$ particles, see
Fig.~\ref{Fs-Earthquaks-N500-N4000}(b).
The data in Fig.~\ref{Fs-Earthquaks-N500-N4000} 
show that run-to-run fluctuations of $\tilde{f_s}$ become
larger when $N$ is smaller, as expected. 
However, we emphasize that the nature of the dynamic fluctuations seems to change qualitatively when the size is
reduced. For a single realization of a quench with a small
enough number of particles, the decay
of the self-intermediate scattering function in the long-time
regime is highly intermittent, as shown by sudden `drops' 
of $\tilde{f_s}$~\cite{kob-barrat2}.
From one run to another, both the temporal location
and the amplitude of these drops  vary considerably, indicating
that in each run the dynamics proceeds via temporally
intermittent relaxation events that mobilize a finite fraction of
the whole system. However, the comparison
with the fluctuations obtained with $N=4000$ suggests that the
spatial extension of these intermittent events does not
increase with the system size, but remains instead
spatially localized.

\begin{figure}
\center
\includegraphics[width=0.95\linewidth]{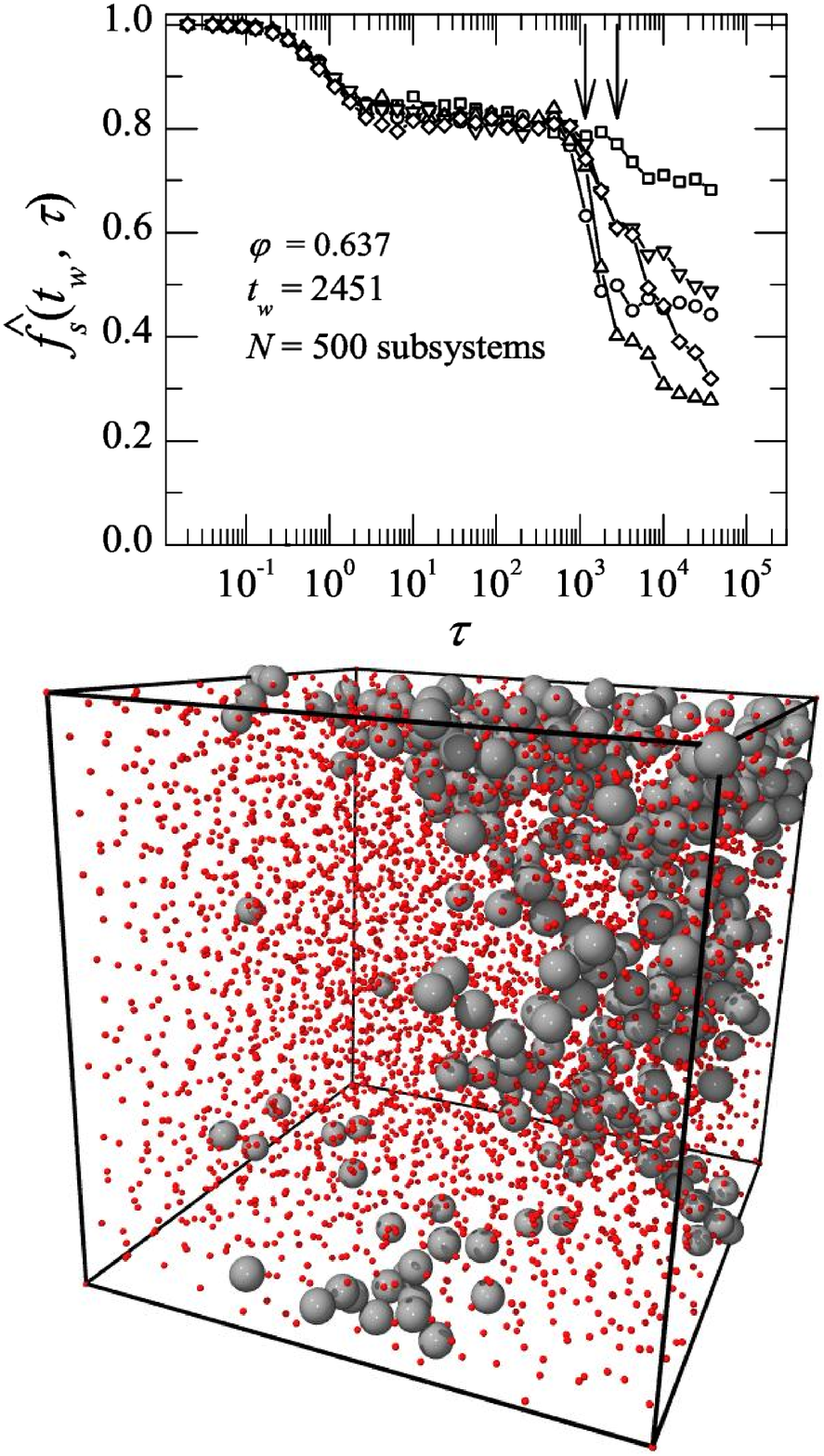}
\caption{Top: intermediate scattering functions $\tilde{f_s} 
(\tw = 2451, \tau)$ measured in sub-systems containing $N=500$ particles
within a larger $N=4000$ system. The variability between
sub-systems in a single realization demonstrates that intermittent
decorrelation events are spatially localized. 
Bottom: snapshot of the $N=4000$ system showing 
the 5\% most mobile particles over the time interval shown by 
the two arrows in the top panel ($\tau = 1174$ and $2789$, respectively) 
are drawn with a larger size. They occupy essentially a small corner 
of the simulation box.}
\label{Fs-Earthquaks-N4000-SmallBoxes}
\end{figure}

Note that this finding could be due to a spurious
finite size effect. An alternative interpretation of the
data in Fig.~\ref{Fs-Earthquaks-N500-N4000} could be that sudden
decorrelations in the system only occur when the system
is too small, and these events are not observed in a bigger system
because they disappear altogether. To disprove this interpretation, we
reanalyze a single quench with $N=4000$ particles, and
further decompose the computation of the correlation function
in 8 distinct sub-systems, each comprising 500 particles and corresponding to a cubic sub-box of linear size $L/2$.
In  Fig.~\ref{Fs-Earthquaks-N4000-SmallBoxes}, we present
the result of this analysis. Remarkably, we find that in
a single quench, a small part of the system might indeed undergo the
type of sudden rearrangement seen in $N=500$ particles runs,
but decorrelation events are not necessarily seen in other
subsystems. This is visible in the top graph, where representative 
$\tilde{f_s}$ for various sub-boxes are shown, and in the bottom snapshot, where the $5\%$ most mobile particles between the two times shown by the arrows in the main plot are depicted as large spheres. Clearly, individual rearrangement events are confined to a fraction of the total volume
(in that case a corner of the box), 
thus affecting significantly only the correlation function of the corresponding sub-box. We conclude that in a large system
the aging dynamics occurs through an intermittent succession of spatially
localized decorrelation events, with no obvious `confinement' effect
imposed by the use of too small a system size, at least
for the two system sizes used in this study.

\begin{figure}
\center
\includegraphics[width=0.95\linewidth]{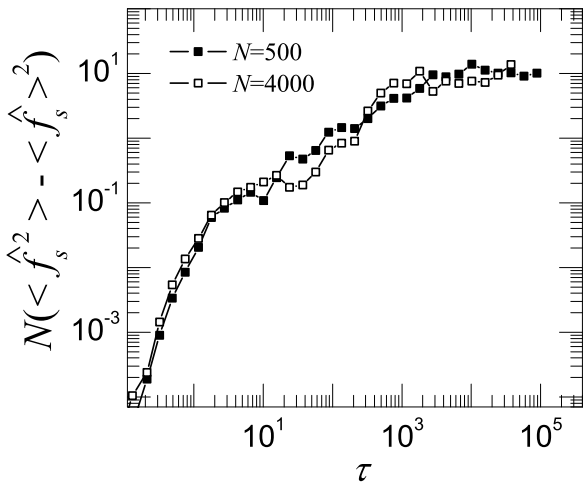}
\caption{The four-point susceptibility $\chi_4$, Eq.~(\ref{chi4}),
for $\tw=2541$, $\phi=0.637$, and two system sizes, shows no
system size dependence, and its modest peak value,
$\chi_4 \approx 10$ confirms the localized nature
of decorrelation events.}
\label{Fluctuations}
\end{figure}

To confirm quantitatively 
both the absence of serious finite size effects, and the
spatially localized nature of decorrelation events, we have measured
the variance of the spontaneous fluctuations of the self-intermediate
scattering function, also known as a `four-point dynamic
susceptibility~\cite{parisi,castillo}:
\be
\chi_4(\tw,\tau) = N \left( \langle \hat{f_s}^2(\tw,\tau) \rangle -
\langle \hat{f_s}(\tw,\tau) \rangle^2  \right) ,
\label{chi4}
\ee
where $\hat{f_s}$ represents the instantaneous
value of $f_s$.
With this normalization, $\chi_4$ should become independent of
the system size in the thermodynamic limit, with an amplitude
that gives direct access to the typical number of particles
relaxing in a correlated manner during aging~\cite{TWBBB,dalle}.
The data for $N=500$ and $N=4000$ shown in
Fig.~\ref{Fluctuations} are essentially the same, within our
statistical accuracy, for both system sizes, and $\chi_4$
reaches a value near $\chi_4 \approx 10$ for timescales
corresponding to the slow relaxation of the correlation function,
confirming the spatially localized nature of relaxation events.

\section{Discussion}
\label{conclusion}

We have studied numerically the aging dynamics
of a concentrated system of nearly hard spheres over a broad
range of volume fractions, and a large time window. We have
addressed the three central questions presented in the
introduction, concerning the evolution of the dynamics,
single particle motion, and the heterogeneous nature of the dynamics.

We found that over the 6 decades of aging times
covered by our simulations,
the structural relaxation continuously slows down,
as expected for any aging system, but we showed that
an asymptotic aging regime is only accessed after
a timescale $\tau_a$, which can become quite large for dense
systems. Interestingly, we found that the relaxation time
$\tau_\alpha(\tw)$ first increases exponentially, as observed
in some experiments on colloidal gels and repulsive Laponite platelets~\cite{lucaaging1,bonn,kaloun}, before crossing over to a
simple aging form, $\tau_\alpha \sim \tw$, at large times,
$\tw \gg \tau_a$.
In the initial regime, we also found that the energy
density decays very quickly, suggesting that the system
evolves rapidly from its highly disordered initial state
to form a nearly frozen structure, which then ages more slowly.
Thus, this initial regime is likely quite sensitive to the detailed
initialization procedure of the system both in simulations
and experiments. The `universal' features of the
aging dynamics of concentrated hard spheres should be discussed
for the second regime only, and our simulations indicate that
concentrated hard spheres follow a simple aging form, as indeed found for
very many glassy materials~\cite{young}.

In the aging regime, we found that particle motions are very restrained,
with particles moving on average less than their own diameter over the
entire duration of simulation. In the aging regime,
$\tw \gg \tau_a$, we found that the self-intermediate functions decay
algebraically at long times, in agreement with several
simulations of soft and hard particle systems~\cite{kob-barrat,puertas}.
This is also  in agreement with recent experiments performed on
colloidal suspensions that are dense enough, so that the aging
regime does not cross over at long times towards
equilibrium behavior~\cite{djamel,martinez}.

We also found that single particle motion
is sub-diffusive, $\Delta r^2(\tw,\tau) \sim \tau^\nu$, with $\nu<1$. This behavior is actually quite consistent with
the results obtained by optical and confocal microsopy on hard and soft colloids, where diffusive behavior is not reached in the experimental time
window~\cite{weeksaging,weeksaging2,yodh}.
This result is also in broad agreement with earlier numerical
work~\cite{kob-barrat},
although subdiffusion was never described in detail before.
We have suggested that a natural
theoretical framework to analyze our results should the
Aging Continuous Time Random Walk (ACTRW), i.e. the
nonequilibrium extension of the random walk picture used
to describe single particle motion at thermal equilibrium
near the glass transition~\cite{pinaki,epl}.
This formalism makes a number of detailed
predictions for the aging regime.
Here, for lack of statistics, we have simply been able to show
that our data
are in qualitatively good agreement with this picture, which can
link the shape of the distribution of particle displacements to
the subdiffusion exponent.
It would be extremely interesting to extend this analysis
in future work and check in more detail how far the
ACTRW picture can be pushed to describe the aging
dynamics of concentrated hard spheres.

We finally showed, by resolving the measurement
of time correlation functions in space and time,
that the aging dynamics of concentrated hard spheres is highly
intermittent, with very sudden relaxation events separating long periods
of time where very little dynamics occurs.
Similar dynamic events have been observed in numerical work before, and were
coined `earthquakes'~\cite{kob-barrat2} or `avalanches'~\cite{lee2}.
Although these names
and additional numerical evidence suggested that these events could well be
system-spanning~\cite{lee2}, we showed by using much larger system
sizes that these events remained actually spatially localized
and do not grow with system size. This is in fact quite consistent with the
experimental report of dynamic clusters that grow
rather modestly in aging colloidal samples~\cite{weeksaging2,yodh},
and with numerical
work reporting a slow growth of four-point dynamic
susceptibilities in aging Lennard-Jones glasses~\cite{castillo}.

Thus, we find that in concentrated hard spheres
the dynamics are intermittent as it occurs in several more complex
colloidal materials, but this leads neither to
anomalous compressed exponential relaxation for time correlation
functions~\cite{lucaaging1},
nor to ballistic motion over large 
distances~\cite{lucaaging1,DuriPRL2009,MazoyerPRL2006,MazoyerPRE2009,maccarrone},
whose origin thus remains largely mysterious. This means that model systems
such as hard spheres or Lennard-Jones glasses are not
good starting points to gain insight into the nature
of these intriguing aging dynamics. It is not clear
what ingredient these models are missing, since similar anomalous
aging dynamics was recently reported in simple `hard' molecular glasses
approaching the glass transition~\cite{CaronnaPRL2008,bob2}. It would be very interesting
to discover a model system displaying the same type of anomalous
aging dynamics, with a particle-scale dynamics that can be
followed in the simulations in the way similar to what we did 
for hard spheres.

\begin{acknowledgments}
This work was financed by Grants NWO-SRON, ANR Dynhet, R\'egion
Languedoc-Roussillon `Chercheurs d'avenir', 
and ACI Jeunes Chercheurs.
\end{acknowledgments}


\begin{thebibliography}{99}

\bibitem{young}
{\it Spin glasses and random fields}, Ed.: A. P. Young (World Scientific,
Singapore, 1998).

\bibitem{leszouches}
{\it Slow relaxations and nonequilibrium
dynamics in condensed matter},
Eds: J.-L. Barrat, J. Dalibard, M. Feigelman, J. Kurchan
(Springer, Berlin, 2003).

\bibitem{lucareview}
L. Cipelletti and L. Ramos,
J. Phys.: Condens. Matter {\bf 17}, R253 (2005).

\bibitem{ediger}
M. D. Ediger, Annu. Rev. Phys. Chem. {\bf 51}, 99 (2000).

\bibitem{naturepusey}
P. N. Pusey and  W. van Megen, Nature {\bf 320}, 340 (1986).

\bibitem{ericluca}
E. Weeks and L. Cipelletti,
to be published.

\bibitem{weeks}
E. Weeks, J. C. Crocker, A. C. Levitt, A. Schofield,
and D. A. Weitz, Science {\bf 287}, 627 (2000).

\bibitem{KegelScience2000} W. K. Kegel and A. van Blaaderen, \textit{Science},  \textbf{287}, 290, 2000.

\bibitem{martinez}
V. A. Martinez, G. Bryant, and W. van Megen,
Phys. Rev. Lett. {\bf 101}, 135702 (2008).

\bibitem{djamel}
D. El Masri, G. Brambilla, M. Pierno, G. Petekidis, A. Schofield,
L. Berthier, and L. Cipelletti,
J. Stat. Mech. P07015 (2009).

\bibitem{SimeonovaPRL2004} N. B. Simeonova and W. K. Kegel, Phys. Rev.
Lett. \textbf{93}, 035701 (2004).

\bibitem{weeksaging}
R. E. Courtland and E. R. Weeks,
J. Phys.: Condens. Matter {\bf 15}, S359 (2003).

\bibitem{weeksaging2}
J. M. Lynch, G. C. Cianci, and E. R. Weeks,
Phys. Rev. E {\bf 78}, 031410 (2008).

\bibitem{lucaaging1}
L. Cipelletti, S. Manley, R. C. Ball, and D. A. Weitz,
Phys. Rev. Lett. {\bf 84}, 2275 (2000).

\bibitem{DuriEPL2006} A. Duri and L. Cipelletti, Europhys. Lett.,  \textbf{76}, 972, 2006.

\bibitem{DuriPRL2009} A. Duri, D. A. Sessoms, V. Trappe and L. Cipelletti, Phys. Rev. Lett.,  \textbf{102}, 085702, 2009.

\bibitem{yodh}
P. Yunker, Z. Zhang, K. B. Aptowicz, and A. G. Yodh,
Phys. Rev. Lett. {\bf 103}, 115701 (2009).


\bibitem{MazoyerPRL2006} S. Mazoyer, L. Cipelletti and L. Ramos, Phys. Rev. Lett.,  \textbf{97}, 238301, 2006.

\bibitem{MazoyerPRE2009} S. Mazoyer, L. Cipelletti and L. Ramos, Phys. Rev. E,  \textbf{79}, 011501, 2009.

\bibitem{bonn}
B. Abou, D. Bonn, and J. Meunier,
Phys. Rev. E {\bf 64}, 021510 (2001).

\bibitem{kaloun}
S. Kaloun, R. Skouri, M. Skouri, J. P. Munch, and F. Schosseler,
Phys, Rev. E {\bf 72}, 011403 (2005).

\bibitem{bob1}
R. Bandyopadhyay, D. Liang, H. Yardimci, D. A. Sessoms,
M. A. Borthwick, S. G. J. Mochrie, J. L. Harden and R. L. Leheny,
Phys. Rev. Lett. {\bf 93}, 228302 (2004).

\bibitem{joshi}
Y. M. Joshi and G. R. K. Reddy,
Phys. Rev. E {\bf 77}, 021501 (2008).

\bibitem{lequeux}
A. Mamane, C. Fretigny, F. Lequeux, and L. Talini, EPL {\bf 5},
58002 (2009).

\bibitem{kob-barrat}
W. Kob and J.-L. Barrat, Phys. Rev. Lett. {\bf 78}, 4581 (1997).

\bibitem{kob-barrat2}
W. Kob and J.-L. Barrat, Eur. Phys. J. B {\bf 13}, 319 (2000).

\bibitem{puertas}
A. M. Puertas,
J. Phys.: Condens. Matter {\bf 22}, 104121 (2010).

\bibitem{CaronnaPRL2008} C. Caronna, Y. Chushkin, A. Madsen and A. Cupane, Phys. Rev. Lett.,  \textbf{1}, 055702, 2008.

\bibitem{bob2}
H. Y. Guo, G. Bourret, M. K. Corbierre, S. Rucareanu,
R. B. Lenox, K. Laaziri, L. Piche, M. Sutton, J. L. Harden,
and R. L. Leheny, Phys. Rev. Lett. {\bf 102}, 075702 (2009).

\bibitem{parisi}
G. Parisi, J. Chem. Phys. B {\bf 103}, 4128 (1999).

\bibitem{castillo}
A. Parsaeian and H. E. Castillo,
Phys. Rev. E {\bf 78}, 060105(R) (2008).

\bibitem{lee1}
K. Vollmayer-Lee, W. Kob, K. Binder and A. Zippelius,
J. Chem. Phys. {\bf 116}, 5158 (2002).

\bibitem{lee2}
K. Vollmayer-Lee and E. A. Baker,
Europhys. Lett. {\bf 76}, 1130 (2006).

\bibitem{buisson}
L. Buisson, L. Bellon, and
S. Ciliberto, J. Phys.: Condens. Matter {\bf 15}, S1163 (2003).

\bibitem{Voigtmann} T. Voigtmann, A. M. Puertas, and M. Fuchs,
Phys. Rev. E {\bf 70}, 061506 (2004).

\bibitem{allen}
M. Allen and D. Tildesley, {\it Computer Simulation of
 Liquids} (Oxford University Press, Oxford, 1987).

\bibitem{brambilla}
G. Brambilla, D. El Masri, M. Pierno, G. Petekidis, A. B. Schofield,
L. Berthier, and L. Cipelletti, Phys. Rev. Lett. {\bf 102}, 085703 (2009).

\bibitem{lee3}
K. Vollmayr-Lee, J. A. Roman, and J. Horbach,
arXiv:1001.0715.

\bibitem{osuji}
A. S. Negi and C. O. Osuji, 
Phys. Rev. E {\bf 80}, 010404 (2009).

\bibitem{struik}
L. C. E. Struik, {\it Physical aging
in amorphous polymers and other materials} (Elsevier, Amsterdam, 1978).

\bibitem{rieger}
U. M\"ussel and H. Rieger,
Phys. Rev. Lett. {\bf 81}, 930 (1998).

\bibitem{kob-barrat3}
W. Kob and J.-L. Barrat,
Phys. Rev. Lett. {\bf 81}, 931 (1998).

\bibitem{pinaki}
P. Chaudhuri, L. Berthier, and W. Kob,
Phys. Rev. Lett. {\bf 99}, 060604 (2007).

\bibitem{epl}
L. Berthier, D. Chandler, and J. P. Garrahan,
Europhys. Lett. {\bf 69}, 320 (2005).

\bibitem{pinaki2}
P. Chaudhuri, Y. Gao, L. Berthier, M. Kilfoil, and W. Kob,
J. Phys.: Condens. Matter {\bf 20}, 244126 (2008).

\bibitem{MW} E. W. Montroll and G. H. Weiss, 
J. Math. Phys. {\bf 6}, 167 (1965).

\bibitem{actrw}
E. Barkai and Y. Cheng, J. Chem. Phys. {\bf 118}, 6167 (2003).

\bibitem{jptrap}
J. P. Bouchaud, J. Phys. I France {\bf 2}, 1705 (1992).

\bibitem{metzler}
R. Metzler and J. Klafter, Phys. Rep. {\bf 339}, 1 (2000).

\bibitem{rottler}
M. Warren and J. Rottler, EPL {\bf 88}, 58005 (2009).

\bibitem{monthus}
C. Monthus and J. P. Bouchaud, J. Phys. A {\bf 29}, 3847 (1996).

\bibitem{TWBBB}
C. Toninelli, M. Wyart, L. Berthier, G. Biroli, and J.-P. Bouchaud,
Phys. Rev. E {\bf 71}, 041505 (2005).

\bibitem{dalle} Dalle-Ferrier
C. Dalle-Ferrier, C. Thibierge, C. Alba-Simionesco, L. Berthier, 
G. Biroli, J.-P. Bouchaud, F. Ladieu, D. L'H\^ote, and G. Tarjus
Phys. Rev. E {\bf 76}, 041510 (2007). 

\bibitem{maccarrone} 
S. Maccarrone \textit{et al.}, to appear in Soft Matter.

\end{thebibliography}
\end{document}